\documentclass{elsart}
                                                                                                                          
\usepackage{psfig,graphicx,amssymb}
                                                                                                                          
\begin{document}
                                                                                                                          
\def\gapprox{\lower.4ex\hbox{$\;\buildrel >\over{\scriptstyle\sim}\;$}}
\def\lapprox{\lower.4ex\hbox{$\;\buildrel <\over{\scriptstyle\sim}\;$}}

\begin{frontmatter}
                                                                                                                          
                                                                                                                          
                                                                                                                          
\title{High energy neutrinos from fast spinning magnetars}


                                                                                                                          
\author{Qinghuan Luo}
                                                                                                                          
\address{School of Physics, The University of Syndey, Australia}
                                                                                                                          
\begin{abstract}
Fast spinning magnetars are discussed as strong sources of high energy neutrinos. 
Pulsars may be born with a short rotation period of milliseconds with
the magnetic field amplified through dynamo processes up to $\sim 10^{15}-10^{16}\,\rm G$.
As such millisecond magnetars (MSMs) have an enormous spin-down power $\sim
10^{50}\,{\rm erg}\,{\rm s}^{-1}$, they can be potentially a strong, extragalactic 
high-energy neutrino source. Specifically, acceleration of ions and subsequent photomeson 
production within the MSM magnetosphere are considered. As in normal pulsars,
particle acceleration leads to electron-positron pair cascades that 
constrains the acceleration efficiency. The limit on the neutrino power as a fraction 
of the spin-down power is calculated. It is shown that neutrinos produced in the
inner magnetosphere have characteristic energy about a few $\times100$ GeV due to the constraint of 
cooling of charged pions through inverse Compton scattering. TeV neutrinos may be produced
in the outer magnetosphere where ions can be accelerated to much higher energies
and the pion cooling is less severe than in the inner magnetosphere. 
High energy neutrinos can also be produced from interactions between ultra-high
energy protons accelerated in the magnetosphere and a diffuse thermal radiation from
the ejecta or from the interaction region between the MSM wind and remnant shell.
The detectability of neutrinos in the early spin-down phase by the current available 
and planned neutrino detectors is discussed.
\end{abstract}
                                                                                                                          
\end{frontmatter}
                                                                                                                          
                                                                                                                          

\section{Introduction}

Young pulsars are generally considered as an important high-energy neutrino source
(e.g. Berezinsky \& Prilutsky 1978; Sato 1978;
Protheroe, Bednarek, \& Luo 1998; Nagataki 2004). Particles can be accelerated
either in the magnetosphere by a rotation-induced electric field
(e.g. Harding \& Muslimov 1998; Hibschman \& Arons 2001, and references therein),
or in the pulsar wind, e.g. by large amplitude waves (e.g. Gunn \& Ostriker 1971;
Sato 1978; Asseo, Kennel, \& Pellat 1978; Usov 1994; Melatos \& Melrose 1996) or 
magnetic field reconnection (e.g. Lyubarsky \& Kirk 2001),
and interact with thermal photons from the star's surface or protons in the remnant 
to produce high energy neutrinos. There is strong observational
evidence for pulsars with an extremely strong magnetic field exceeding the quantum
electrodynamics (QED) critical field $B\gg B_c\approx 4.4\times10^{13}\,{\rm G}$,
called magnetar (e.g. Woods \& Thompson 2004). Soft gamma-ray repeaters (SGR)
and anomalous X-ray pulsars (AXP) are believed to be slowly rotating magnetars 
(Thompson \& Duncan 1995). The possible existence of fast spinning magnetars has also been 
proposed (e.g. Usov 1994; Blackman \& Yi 1998; Wheeler et al 2000; Rees 
\& M\'esz\'aros 2000; Gaensler et al. 2004). Such MSM may be formed from an accretion-induced collapse, in 
which the magnetic field is amplified exponentially to $10^{15}-10^{16}\, \rm G$ by dynamo amplification 
(Dar et al. 1992; Duncan \& Thompson 1992; Thompson 1994). If such fast spinning 
magnetars exist they are by far the most energetic pulsars with a typical 
spin-down power $10^{49}-10^{50}\,{\rm erg}\,{\rm s}^{-1}$ for a magnetic field 
$B_0=10^{15}\,{\rm G}$ and a rotation period $P=1\,{\rm ms}$, which is 
about $10^{12}$ times more powerful than the Crab pulsar. 
MSMs spin down rapidly, producing powerful transient gamma-ray emission within a 
typical spin-down time $\sim (1/2)I\Omega^2_0/L_E\approx10^3\,\rm s$,
where $\Omega_0$ and $I\approx10^{45}\,{\rm g}\, {\rm cm}^2$ are the initial angular velocity and 
inertial moment of the pulsar, and $L_E$ is the spin-down luminosity due to the magnetic 
dipole radiation. As in normal pulsars, rotation-driven acceleration leads to
gamma-ray emission and in particular, acceleration of ions may lead to photopion 
production through either proton-photon or proton-proton interactions, producing
neutrinos. Thus, MSM can be a potentially observable source of high-energy neutrinos 
during its initial spin-down phase.

Neutrinos originating from magnetospheric acceleration has 
been considered by several authors for normal young pulsars (e.g. Sato 1978;
Protheroe, Bednarek, \& Luo 1998; Bednarek \& Protheroe 1997; Bednarek \& 
Protheroe 2002). Neutrinos from slowly rotating magnetars were discussed 
recently by Zhang et al. (2003). However, as such slow rotators have a much lower 
spin-down power than the Crab pulsar (by a factor of $10^5$), the corresponding 
luminosity of rotation-powered
neutrino emission is marginal even for a km scale neutrino detector.
So far, in most of the magnetospheric-origin models for neutrino emission, the constraints
by pair production and radiation reaction were not considered. Magnetospheric
acceleration arises from a rotation-induced electric field in the open field line region
due to that the charge density of outflowing charged particles deviates
from the corotation charge density, referred to as the Goldreich-Julian (GJ) density.
Pair cascades tend to screen out the accelerating electric field and 
send a backflow of opposite-charged particles that can reverse
the sign of the electric field. Therefore, pair cascades can 
strongly limit the fraction of the spin-down power going into particle 
acceleration and constrain the neutrino luminosity. 
There are extensive discussions on neutrino
emission resulting from particle acceleration in the pulsar wind or at the wind
termination shock (e.g. Berezinsky \& Prilutsky 1978; Sato 1978; Beall \& Bednarek
2002; Granot \& Guetta 2003; Nagataki 2004; Luo 2005).
In these models, it is generally hypothesized that protons (or ions) can be 
accelerated efficiently to ultra-high energy. Although acceleration in the 
pulsar wind can be efficient in the sense that most of the Poynting flux that is
thought to be dominant near the pulsar is converted to particle kinetic energy, 
as supported by the observational evidence from the Crab nebula (Gaensler et al.
2002), the specific acceleration mechanism is poorly understood (e.g. Kirk 2005) and
there is no reliable estimate of the maximum energy of accelerated protons. 

In this paper, we explore the magnetospheric origin of neutrinos from MSMs in which, 
compared to normal young pulsars, acceleration processes are strongly 
modified by both the supercritical magnetic field and rapid rotation. 
There are basically two classes of model for particle acceleration in
the pulsar magnetosphere: the polar gap model, in which acceleration is
assumed to occur near the polar cap (PC), and the outer gap model, in
which the acceleration region is located in the outer magnetosphere.
There is also a variant between the two, called the slot gap model
(e.g. Arons 1983; Muslimov \& Harding  2004). We consider these two cases 
separately and comment briefly on the slot gap model in Sec. 6. 
For acceleration near the PC we extend the polar gap model for normal pulsars
(e.g. Hibschman \& Arons 2001; Harding \& Muslimov 1998) to MSMs. Since
the steady gap model may not be realistic as pair cascades are likely nonstationary and
time-dependent, we emphasize the energetics of polar gap acceleration rather than 
a specific gap model. The energetics, described by the acceleration efficiency, 
which is defined here as the ratio of pair-production limited 
potential to the maximum potential (across the PC), determining 
how much of the spin-down power goes into particle acceleration, can be determined 
from the relevant free path for pair production. A pair may be produced through interaction
of a Lorentz boostered thermal photon with the Coulomb field of an ultrarelativistic ions.
The more efficient pair production process is single photon decay in a strong magnetic
field where energetic photons are emitted by electrons or positrons.
In the supercritical magnetic field, the probability for a single photon decay into a 
pair can be approximated by a step function and thus, the free path corresponds to that 
for the pair production threshold. 

Since the outer gap is generally considered more
efficient than the inner gap in conversion of the spin-down power to
particle acceleration, we re-examine the possibility of neutrino production
due to acceleration in the outer magnetosphere. Ion acceleration in the 
outer vacuum gap, as a source of high energy neutrinos, 
was discussed long ago by Sato (1978) and recently by Bednarek \& Protheroe
(1997) and, as high energy cosmic ray sources, was 
considered by Bednarek \& Protheroe (2002). In those
cases, the detailed mechanism of ion injection and constraints by pair production on
the acceleration were not discussed. In contrast to the inner gap scenario,
acceleration of ions in the outer gap, which is detached from the star's
surface, requires a specific injection mechanism. Furthermore, the usual vacuum gap model 
(e.g. Cheng et al 1986) does not permit an external particle flux flowing through the gap.
The effect of such external flux on the outer gap was considered recently
by Hirotani \& Shibata (1999), who showed that the gap location can be strongly
modified due to injection of particles into the gap. 
Here, we consider the possibility that ion injection is due to 
a global current that circulate the pulsar-wind system (e.g. Shibata 1991;
Hirotani \& Shibata 1999) and that these ions are accelerated to ultra-high 
energies in the outer gap, leading to photomeson processes that 
produce high energy neutrinos. As for the polar gap scenario, one 
concentrates on the energetics and the particle flux, which can be determined
respectively by the radiation-reaction limit and the relevant free path 
for pair production. The result should not depend much on the specific
detail of the outer gap model.
 
In Sec. 2, pair production processes in the MSM and constraints on
the acceleration efficiency near the PC are considered. Photomeson production 
on thermal radiation from the neutron star's surface is discussed in Sec. 3. 
The relevant neutrino flux originating from the inner magnetosphere is 
estimated and its detectability is discussed. In Sec. 4, proton acceleration 
in the outer gap is discussed. Photopion production in the outer magnetospheric region
is considered in Sec. 5. The conclusions are summarized in Sec. 6. 

\section{Acceleration of ions near the PC}

A neutron star forms with a hot surface on which free emission of charged particles 
occurs. We assume that for pulsars with $\mbox{\boldmath $\mu$}_m\cdot{\bf \Omega}<0$ (where 
$\mbox{\boldmath $\mu$}_m$ is the magnetic pole and $\bf \Omega$ is the angular velocity), 
free emission of ions (nuclei) from the PC occurs. These ions are accelerated along the field lines to 
ultra-high energies. Thermal photons from the star's hot surface can be Lorentz
boostered to ultrarelativistic energy in the nucleus rest frame and
produce pairs in the nucleus Coulomb field. Some electrons will be accelerated downward 
to initiate a cascade producing more pairs near the star, which provides seed positrons. 
These seed positrons are accelerated outward, initiating a further cascade above the PC. 
Whether the constraint on acceleration is determined by the proton free path or 
positron free path depends on nature of the polar gap. If the gap is nonstationary,
the cascade by the seed positrons may play the more important role and if the acceleration 
proceeds in a steady, time-independent manner, the constraint is provided by protons. 
Both cases are considered here with particular emphasis on the former.

\subsection{Accelerating potential}

The maximum available potential across the PC can be written as
\begin{equation}
\phi_{\rm m}={\textstyle{1\over2}}\theta^4_dB_0R_0\approx 6.6\times10^{21}
\Biggl({B_0\over10^{15}\,{\rm G}}\Biggr)
\Biggl({1\,{\rm ms}\over P}\Biggr)^2\, {\rm V},
\label{eq:fim}
\end{equation}
where $B_0$ is the magnetic field on the PC,
$\theta_d=(2\pi R_0/cP)^{1/2}\approx0.46(1\,{\rm ms}/P)^{1/2}$ is the half-opening 
angle of the PC, $R_0=10^6\,\rm cm$ is the star's radius. The specific form 
of the potential depends on the detail of a particular model (e.g. 
Arons 1983; Harding \& Muslimov 1998; Shibata et al. 1998). The basic assumption that
is commonly used is that a nearly-parallel electric field develops due to the deviation of
the charge number density of the outflow from the GJ density $n_{GJ}\approx
B/2ZecP$, where $Ze$ is the charge of the particle. In the case of outflowing ions, since
its composition is not well constrained by observations (Lai 2001), 
protons are considered.

For acceleration sufficiently near the PC, the potential can be assumed to be a 
simple quadratic function of the distance from the PC 
(in units of $R_0$), \begin{equation}
\phi=\phi_{\rm m}\epsilon s^2,
\label{eq:fim1}
\end{equation}
where we consider the region $s<\theta_d$ and $\epsilon$ is the parameter that 
incorporates geometric effects such as the 
inclination angle between the rotation axis and the magnetic pole and the 
latitudes of the open field lines, the general relativity effects such as the frame-dragging
(e.g. Harding \& Muslimov 1998, 2001; Hibschman \& Arons 2001), and 
the detail how the low-altitude potential is matched to that at higher altitudes. 
For example, for a vacuum gap one has $\epsilon\sim1/\theta^2_d$, which implies that the
maximum potential is achieved at $s\sim \theta_d$ (Ruderman \& Sutherland
1975). If the potential is due to the space-charge effect, it increases with the distance
more slowly than in the vacuum gap, with  $\epsilon\sim 1/\theta_d$ if the frame-dragging
is dominant (e.g. Harding \& Muslimov 2001; Hibschman \& Arons 2001) and 
$\epsilon\sim 1$ if the inclination angle is
nearly orthogonal (e.g. Harding \& Muslimov 2001).
Let $\phi_f=\phi_{\rm m}\epsilon s^2_f$, where $s_f$ is the 
characteristic distance where one pair is produced per primary particle. The 
efficiency can be estimated as 
\begin{equation}
\eta={\phi_f\over\phi_{\rm m}}\approx \epsilon s^2_f,
\end{equation}
provided that the energy loss is not dominant over the acceleration.
It can be shown that only a very small fraction of the spin-down power can be
extracted through particle acceleration in the magnetosphere of a young 
magnetar.

\subsection{The acceleration efficiency limited by pair production}

To estimate the acceleration efficiency limited by pair production, 
one derives the characteristic distance at which one pair is produced per 
primary protons or positrons (or electrons). First consider
pair production by protons. Ultrarelativistic protons can 
produce pairs on a thermal radiation field through interaction of thermal photons in 
the proton's Coulomb field at a rate
\begin{equation}
{dN\over ds}\approx \sigma_{p\gamma}n_{\rm ph}R_0,
\end{equation}
where $\sigma_{p\gamma}\approx(3\alpha_f\sigma_T/8\pi)\ln(4\varepsilon_{th}\gamma)
\approx 5\times10^{-27}\,{\rm cm}^2$ is the pair production cross section due to
proton-photon collision (Chodorowski, Zdziarski, \& Sikora 1992), 
$\alpha_f\approx1/137$ is the fine constant,
$\sigma_T\approx6.7\times10^{-25}\,{\rm cm}^2$ is the 
Thomson cross section, and $\varepsilon_{th}=2.8\Theta$ is the thermal photon energy
in terms of the dimensionless surface temperature $\Theta\equiv k_BT_s/m_ec^2=
8.4\times10^{-3}\,(T_s/5\times10^7\,{\rm K})$.  We assume 
$2\varepsilon_{th}\gamma\gg1$. The photon number density at a radial 
distance $r$ is 
\begin{eqnarray}
n_{\rm ph}&\approx&0.36{\sigma_{SB}T^3_s\over ck_B}\,\left({R_0\over
r}\right)^2\nonumber\\
&\approx& 6.2\times10^{23}\left({R_0\over r}\right)^2
\left({T_s\over 5\times10^7\,{\rm K}}\right)^3\,{\rm cm}^{-3},
\label{eq:nph1}
\end{eqnarray}
where $\sigma_{SB}$ is the Stefan-Boltzmann constant. The proton free path can be estimated from 
\begin{equation}
s_f\approx {1\over\sigma_{p\gamma}n_{\rm ph}R_0}.
\label{eq:proton}
\end{equation}
One estimates $s_f\sim 3.2\times10^{-4}$ for $T_s=5\times10^7\,\rm K$, which
increases slowly with an increasing pulse period $P$ and rapidly with a decreasing
temperature $T_s$. The proton free path is generally longer than
that for an electron (or a positron) through single photon decay in 
a strong magnetic field. The latter is derived below.

In deriving the electron (or positron) free path, it is worth noting 
two important features of pair production through a
single photon decay in a supercritical magnetic field $B\gg B_c$. First,
the probability for single photon decay into a pair can be 
approximated by the step function, i.e.  one pair is produced when the photon 
energy reaches the threshold. Second, in the supercritical magnetic field, pairs are
produced mostly in the ground state (Daugherty \& Harding 1983; Weise \& 
Melrose 2002), which tends to suppress further pair creation.
The second feature implies that due to the absence of synchrotron photons
from the secondaries the pair cascade is less efficient than in moderately strong
magnetic fields ($B\ll B_c$). The first feature implies that estimate of the free 
path is reduced to calculation of the characteristic path for 
the pair creation threshold, given by (Harding, Baring \& Gonthier 1997)
\begin{eqnarray}
\varepsilon_\gamma&>&{2\over\sin\theta_{kB}},\quad {\rm for\ the}\ \Vert\ {\rm photon},
\label{eq:threshold1}\\
\varepsilon_\gamma&>&{1+(1+2\varepsilon_B)^{1/2}\over\sin\theta_{kB}},\quad 
{\rm for\ the}\ \perp\ {\rm photon},
\label{eq:threshold2}
\end{eqnarray}
where all relevant energies are dimensionless in units of $m_ec^2$,
$\varepsilon_B=B/B_c$ is the cyclotron energy, 
the $\perp$ and $\Vert$ modes correspond respectively to photons with polarization 
perpendicular to and in the plane defined by the wave vector $\bf k$ and the magnetic field 
$\bf B$, and $\theta_{kB}$ is the propagation angle (with respect to $\bf B$).
Note that the threshold for the $\perp$ photon is higher than that for the $\Vert$ photon.
Since in the superstrong magnetic field the $\perp$ photons can be efficiently split into 
the $\Vert$ photons, only the first threshold condition is relevant here. 

Two main radiation processes considered here that can produce pairs are 
resonant inverse Compton scattering (RICS) (Herold 1979; Gonthier 
et al. 2000) and curvature radiation.
RICS occurs at $\varepsilon_B\sim \varepsilon_{th}\gamma$ in the rest frame.
The energy of the RICS photon emitted by an electron (or positron) is 
$\varepsilon_s\sim \varepsilon_B\gamma$ (in the observer's frame). The $\Vert$ 
photon emitted at $s_i$ will be absorbed and convert to a pair at $s_f$, provided that 
\begin{equation}
s_f-s_i\geq{a_c\over\gamma\varepsilon_B},
\label{eq:threshold3}
\end{equation}
where $a_c=4/3\theta_*\geq4/3\theta_d$ is the curvature radius (in $R_0$) of 
the field line with the magnetic colatitude $\theta_*$ on the PC. 
One has $a_c=4/3\theta_d\approx2.9(P/1\,{\rm ms})^{1/2}$ for
the last open field lines.
The condition (\ref{eq:threshold3}) can be derived directly from (\ref{eq:threshold1}).
For $B\gg B_c$, the effective cross section for resonance scattering can be written as 
$\sigma_{\rm eff}=(3\pi\sigma_T/4\alpha_f)f_B$, 
where $f_B$ describes the relativistic effect that leads to the suppression of
the cross section in the superstrong magnetic field. 
The photon production rate can be estimated from
$dN_s/ds\approx N_m(\gamma_m/\gamma)^2$,
where $N_m=[9x\Theta\varepsilon_B/(8\pi^2\gamma^2_m)]( \sigma_{\rm eff}R_0/\lambda^3_c)$, 
$\lambda_c=\hbar/m_ec$ is the Compton wavelength, 
$x=-\ln[1-\exp(-\varepsilon_B/\Theta
\gamma(1-\beta\cos\theta_m))]$, $\theta_m$ is the
maximum propagation angle of the incoming photon. For thermal radiation from 
the whole surface, one has $\cos\theta_m=[(s+1)^2-1]^{1/2}/(s+1)$. For 
$B_0=10^{15}\,\rm G$, one may assume $f_B\approx0.01$ (Gonthier et al. 2000). 
The photon production rate depends strongly on the distance $s$ through 
$\gamma=\gamma_m\epsilon s^2$, where 
$\gamma_m=e\phi_m/m_ec^2\approx 1.3\times10^{16}(1\,{\rm ms}/P)^2(B_0/10^{15}\,{\rm G})$.
Therefore, one may obtain the following condition for one pair-producing photon to be 
produced per primary:
\begin{equation}
N_s\approx {N_m\over3\epsilon^2}\left({1\over s^3_i}
- {1\over s^3_f}\right)=1.
\label{eq:Ns}
\end{equation}
For $s_i\ll s_f$, one may obtain $s_i\approx (N_m/3\epsilon^2)^{1/3}$
by neglecting $1/s^3_f$ in (\ref{eq:Ns}). From (\ref{eq:threshold3}), one derives
\begin{eqnarray}
s_f&\approx&{a_c
\over \gamma_m\epsilon s^2_i\varepsilon_B}=
{3^{2/3}a_c\epsilon^{1/3}\over \gamma_mN^{2/3}_m \varepsilon_B}\nonumber\\
&\approx&
10^{-3}\epsilon^{1/3}
\left({5\times10^7\,{\rm K}\over T_s}\right)^{2/3}\left({0.5\over x}\right)^{2/3}
\left({B_0
\over 10^{15}\,{\rm G}} \right)^{-4/3}\left({1\,{\rm ms}\over P}\right)^{1/6}.
\label{eq:sf1}
 \end{eqnarray}
The gap length limited by RICS is about 
$s_f\approx10^3\,{\rm cm}/R_0$ for the above nominated parameters.
Since pairs are produced near threshold, in 
the ground state, further generations of pairs have to be
produced by photons from RICS or curvature radiation by the secondary pairs. The
photons from the latter have energy too low to produce pairs.
Since $-E_m\sim 2\phi_m\epsilon s_f/R_0\sim 10^{13}\epsilon\,
{\rm V}\,{\rm cm}^{-1}$ is much higher than that ($2\times10^{10}\,{\rm V}\,{\rm cm}^{-1}$) 
required for ionizing positroniums (Usov \& Melrose 1996), pairs produced
inside the gap are mostly unbound. 
It is worth emphasizing that the characteristic length of a MSM polar gap
is very small, much smaller than the PC radius. Thus, the quadratic 
potential (2) is a good approximation. Since a large backflow of particles may 
occur, the gap 
may be highly nonstationary with a characteristic frequency in the radio band,
$\Omega_c\approx c/s_fR_0\approx 3\times10^7\,{\rm s}^{-1}$. 

Radiation from nonresonant inverse Compton scattering may contribute to pair 
production. It can be shown that this process is less important than RICS.
In the Klein-Nishina regime, which is relevant here, the photon
production rate is $dN_s/dt\sim -\dot{\gamma}/\gamma$, with
$\dot{\gamma}$ the energy loss rate. Since $\gamma\sim s^2$ and $-\dot{\gamma}\approx
(\pi^2\sigma_Tc/120\lambda^3_c)\Theta^2\ln(4\gamma \Theta)\approx
2\times10^{12}(T_s/5\times10^7\,{\rm K})^2\ln(4\gamma \Theta)$,
one may estimate $s_i\sim (-\dot{\gamma}R_0/\gamma_m \epsilon c)$.
From (\ref{eq:threshold1}) one finds $s_f\sim a_c/\gamma_m s^2_i
\sim 1/[\ln(4\gamma_ms^2_i\Theta)]^2$.

Pair production by curvature radiation can compete with RICS in limiting the acceleration.
The power of curvature radiation is 
\begin{equation}
P_{curv}={\textstyle{2\over3}}{ce^2\over R^2_c}\gamma^4,
\label{eq:Pc}
\end{equation}
which gives the characteristic energy  $\varepsilon_c\approx (\lambda_c/R_c)\gamma^3$,
where $R_c=a_cR_0$. The production rate of curvature photons is $P_{curv}/\varepsilon_cm_ec^2$, 
which can be written into the approximate expression $dN_s/ds\approx
(2r_e\gamma_m/3a_c\lambda_c)\epsilon s^2$,
where $r_e=e^2/m_ec^2\approx2.8\times10^{-13}\,\rm cm$ is the classical electron radius.
Assuming at least one photon in the energy $\varepsilon_c$ be emitted through curvature radiation, 
one may obtain $s_f$ from $N_s\approx (2r_e\gamma_m/9a_c\lambda_c)\epsilon\,s^3_f$.  
The condition for $N_s=1$ yields
\begin{equation}
s_f\approx \left({9a_c\lambda_c\over2r_e\gamma_m\epsilon}\right)^{1/3}\approx5\times10^{-5}
\epsilon^{-1/3}\left({B_0\over10^{15}\,{\rm G}}
\right)^{-1/3}\left({P\over1\,{\rm ms}}\right)^{5/6}.
\label{eq:sf2}
\end{equation}
The Lorentz factor of a positron at $s_f$ is
$\gamma_f=\gamma_m\epsilon s^2_f\approx 5\times10^7\epsilon^{1/3}$ for a millisecond
period, which is close to that limited by the energy loss due to 
curvature radiation, given by 
$\gamma_c\approx a^{4/7}_c\gamma^{1/7}\left({3R_0/r_e}\right)^{2/7}
\approx8.2\times10^7$.
The energy of the pair-producing photon is given by
\begin{eqnarray}
\varepsilon_c=\left({\lambda_c\over a_c R_0}\right)\gamma^3
\approx
 3.2\times10^6\epsilon\left({B_0\over10^{15}\,{\rm G}}\right)\left({1\,{\rm ms}\over P}\right)^2.
\end{eqnarray} 
The photon must travel a further distance, $\Delta s$, before decaying into a pair. This distance can be
determined from the threshold condition: $\varepsilon_c=2a_c/\Delta s$.
One estimates $\Delta s\approx 2a_c/\varepsilon_c\approx 10^{-5}\ll s_f$. 
Therefore, pair production due to curvature radiation puts a much more severe 
constraint on the acceleration than RICS. The corresponding efficiency limited by pair production
by curvature radiation can be written as
\begin{equation}
\eta\approx \epsilon\left({9a_c\lambda_c\over2r_e\gamma_m\epsilon}
\right)^{2/3}\approx3\times10^{-9}\, \epsilon^{1/3}
\left({B_0\over 10^{15}\,{\rm G}}\right)^{-2/3}
\left({P\over1\,{\rm ms}}\right)^{5/3}.
\label{eq:eta}
\end{equation}
The efficiency limited by pair production by protons is given by
\begin{equation}
\eta\approx {\epsilon\over(\sigma_{p\gamma}n_{\rm ph}R_0)^2}\sim 10^{-7}\epsilon
\left({T_s\over5\times10^7\,{\rm K}}\right)^{-6}.
\label{eq:eta2}
\end{equation}
Other pair production processes such as pair production by photon-photon collision
may also limit the acceleration near the pulsar. The pair production rate via 
$\gamma\gamma\to e^\pm$, however, is significantly reduced in the superstrong magnetic field 
(Daugherty \& Bussard 1980; Baring \& Harding 1992). Since $\eta\propto \epsilon^{1/3}$, 
the efficiency (\ref{eq:eta}) is not very sensitive to the specific model of the potential
and all the three cases discussed (cf. Eq. \ref{eq:fim1}) lead to a very similar result
except when the period increases substantially (cf. Sec. 2.3).

In general, the efficiency determined by the proton free path is significantly
higher (cf. Eq~\ref{eq:proton} and \ref{eq:eta2}) than that determined by 
pair-production free paths of positrons (electrons). However, the latter is favoured as 
there seems no evidence from radio observations for two different classes of 
pulsars: one class with a proton-controlled gap and the other with an electron/positron-controlled
gap. In realistic situations, the polar gap acceleration is likely to be nonstationary with protons
playing only a minor role in the gap dynamics, and the gap energetics is 
limited by pair production by electrons/positrons.

\begin{figure}
\psfig{file=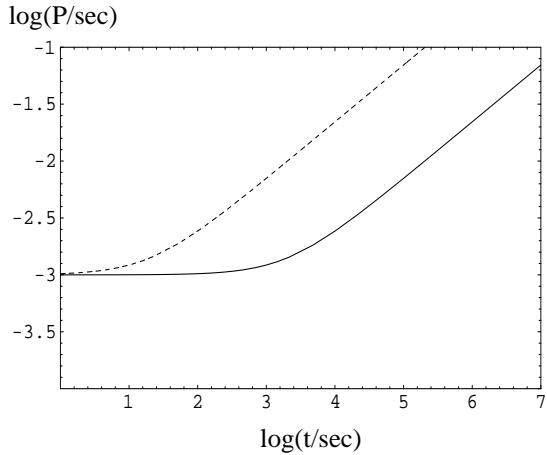,width=8cm}
\caption{Pulsar period as a function of time. The solid and dashed lines correspond
to $B_0=10^{15}\,\rm G$ and $10^{16}\,\rm G$, respectively.}
\label{fig:p}
\end{figure}

\subsection{Evolution of the MSM polar gap}

\begin{figure}
\psfig{file=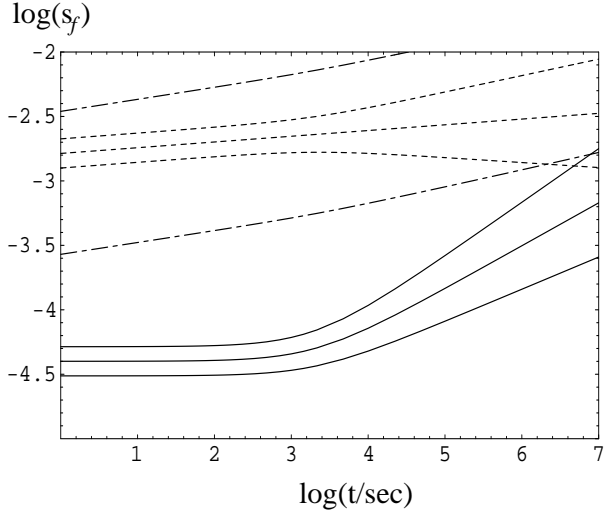,width=8cm}
\caption{The free path for pair creation. The dashed lines is for RICS, which
from top to bottom correspond to $\epsilon=1/\theta^2_d$, $1/\theta_d$ and 1.
The three solid lines is for curvature radiation, corresponding to
$\epsilon=1/\theta^2_d$ (lower), $1/\theta_d$ (middle) and 1 (upper). 
One assumes $T_s\sim 5\times10^7\,\rm K$ and $B_0\sim 10^{15}\,{\rm G}$.
For comparison, the proton's free path is shown as the dash-dot lines for 
$T_s=2\times10^7\,\rm K$ (upper) and $5\times10^7\,\rm K$ (lower). 
For $\epsilon=1$, RICS takes over the effect of curvature emission at about $t\sim 46\,\rm day$
when the period increases to about $P\sim 32\,\rm ms$. For other two cases, RICS
becomes dominant at $t>10^7\,\rm s$ (not shown in the figure). Since $\theta_d\propto 1/P^{1/2}$ and 
$\theta_d\sim 5.5\times10^{-2}$ at $t\sim 10^7\,\rm s$, one always has $s_f\ll\theta_d$ in 
the parameter regime considered here and the quadratic potential (\ref{eq:fim1}) should be
a good approximation. 
}
\label{fig:sf}
\end{figure}

Both $a_c$ and $P$ are time-dependent, which leads to time-dependence of the efficiency.
In the case where pair production by electrons/positrons involves RICS or proton-photon collision, 
the relevant free path depends on the surface temperature, which evolves
with time. To derive the evolution of the MSM polar gap, we assume that the pulsar spin-down is 
due to the magnetic dipole radiation. Spin-down through gravitational waves 
may be important at the initial stage (e.g. Arons 2003) and such complication is not 
considered here. The pulsar period as a function of time is then given by
\begin{equation}
P(t)=P_0\left(1+{t\over \tau_d}
\right)^{1/2},
\label{eq:Pt}
\end{equation}
where $P_0=1\,\rm ms$ is the initial period, and
$\tau_d=3c^3IB^2_0R^6_0(P_0/2\pi)^2\approx 2\times10^3\,\rm s$ for $B_0=10^{15}\,{\rm
G}$ and $I=10^{45}\,{\rm g}\,{\rm cm}^2$
 is the typical spin-down time. The pulsar period $P$ as a function of $t$
is shown in Figure~\ref{fig:p}. 
The free path that defines the gap length is shown
in Figure~\ref{fig:sf} for the cases of curvature radiation (solid line)
and RICS (dashed line). The dash-dot lines represent the proton free path
for $T_s=2\times10^7\,\rm K$ (upper) and $5\times10^7\,\rm K$ (lower). 
Since the spin-down time is much shorter than the cooling
time of the neutron star, decreasing surface temperature does not modify 
much RICS. However, it does affect the proton free path. Since
the initial cooling processes are not well understood, to include this effect 
we model the star's cooling as a simple power-law function of time, i.e.
$T_s(t)\sim T_0(t/1\,{\rm s})^{-\delta}$, with $\delta=1/30$ being used in
in Figure~\ref{fig:sf} for protons and RICS. 

\begin{figure}
\psfig{file=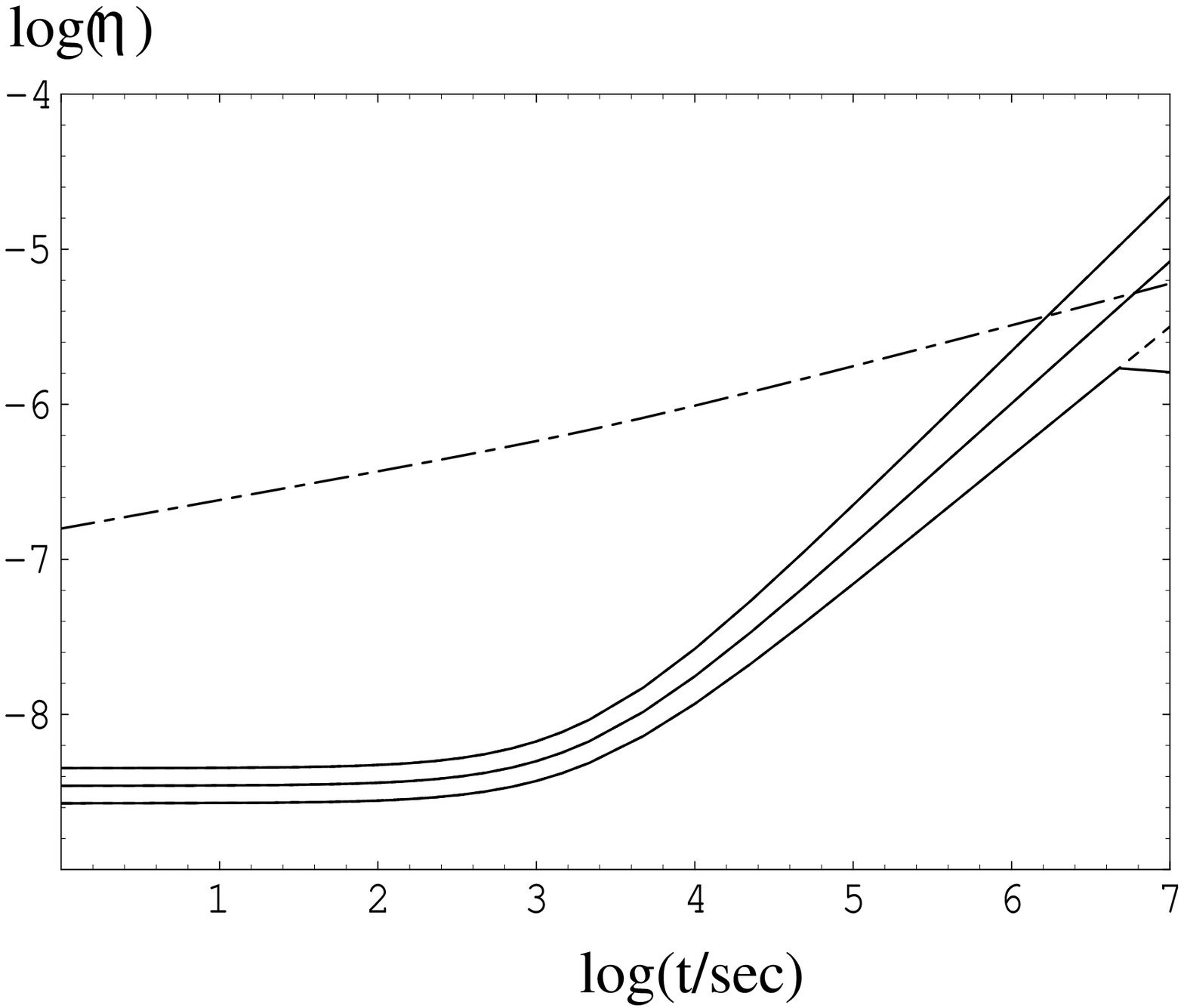,width=8cm}
\caption{The evolution of the efficiency for $T_s(0)=2\times10^7\,\rm K$ (dashed) and
$5\times10^7\,\rm K$ (solid). 
In each case, lines from top to bottom correspond to $\epsilon=1/\theta^2_d$, $1/\theta_d$
and 1. The break corresponds to that pair production due to curvature radiation
switches over to that due to RICS for $\epsilon=1$. For the other
two cases, the breaks occur at a later time (not shown in this figure). 
The dash-dot-dashed line corresponds to the efficiency determined by the proton
free path (for $T_s(0)=5\times10^7\,\rm K$), 
which is significantly higher than that by positrons/electrons.
As in Figure~\ref{fig:sf}, we assume $B_0=10^{15}\,\rm G$.
}
\label{fig:eta}
\end{figure}

For MSMs, the effects of the RICS and curvature radiation
on $s_f$ are comparable at the period
\begin{equation}
P_c\approx 22\epsilon^{2/3}\left({5\times10^7\,{\rm K}\over T_s}\right)^{2/3}
\left({0.5\over x}\right)^{2/3}\left({B_0\over 10^{15}\,{\rm G}}\right)^{-1}\,{\rm ms}.
\label{eq:Pc2}
\end{equation}
This is in contrast with normal young pulsars in which RICS is generally
the more important in constraining the energetics of the polar gap.
Since in a supercritical magnetic field, RICS is suppressed and 
acceleration is extremely rapid (for $P\sim 1\,\rm ms$), pair production due to curvature
radiation is dominant over RICS until the period reaches the period
(\ref{eq:Pc2}). The efficiency $\eta$ as a function of time is shown in 
Figure~\ref{fig:eta}. For $\epsilon\sim1$, this period 
is found to be $P_c\sim 32\,\rm ms$ (cf. Figure~\ref{fig:eta})
when $t\sim 10^{6.6}\,\rm s$ (about 46 day). 
The gap efficiency, initially constrained by pair production due to curvature radiation,
increases as the MSM spins down. When the RICS becomes dominant at $P_c$,
$\eta$ decreases gradually because effective RICS favours moderate acceleration.
Generally, the efficiency decreases to a minimum and then the effect of the surface cooling
takes over, which renders an increase in $\eta$ due to a decreasing $T_s$.
                                                                                     
\section{Photomeson production}

The outward-propagating relativistic protons can produce neutrinos on the
thermal radiation from the star's surface through the photomeson process.
The dominant channel for neutrino production is photomeson 
production at the $\Delta$ resonance (Stecker et al. 1991;
Waxman \& Bahcall 1997), corresponding to the process $\gamma p\to\Delta\to n\pi^+$.
The pion decay, $\pi^+\to \mu^++\nu_\mu\to e^++\nu_e+\bar{\nu}_\mu+\nu_\mu$, produces neutrinos. 
For the photon energy $\varepsilon'_\gamma$ in the rest frame of the proton, the cross section 
is peaked at $\varepsilon'_\gamma\approx \varepsilon_0=0.35\mu_p$ with the width
$\Delta \varepsilon_0\approx0.12\mu_p$ (Stecker et al. 1991). 
The Lorentz factor $\gamma_*=1/(1-\beta^2_*)^{1/2}$ 
of the proton at which the cross section for pion production peaks is estimated to be
\begin{equation}
\gamma_*={\varepsilon_0\over(1-\beta_*\cos\theta)\varepsilon_{th}}\approx
2.7\times10^4\left({5\times10^7\,{\rm K}\over T_s}\right){1\over 1-\beta_*\cos\theta},
  \label{eq:gstar}
\end{equation}
where $\theta$ is the propagation angle of the 
incoming photon, given by $\cos\theta=[(1+s)^2-1]^{1/2}/(1+s)$.
Since one has $\cos\theta\approx (2s)^{1/2}$ for $s\ll1$ near the surface, the 
factor $1/(1-\beta_*\cos\theta)$ can be discarded. Although the proton free path gives
higher $\eta$ than electrons/positrons (cf. Figure~\ref{fig:eta}), 
which may be applicable for a steady gap, we emphasize the latter case 
which is applicable when acceleration is nonstationary (cf. Sec. 2). Assuming 
the characteristic acceleration length is $s_f$, which is limited by 
pair production by seed positrons, one obtains the maximum Lorentz 
factor of the accelerated proton, 
\begin{equation}
\gamma_f={e\phi_{\rm m}\eta\over m_pc^2}={\gamma_{\rm m}\eta\over \mu_p}.
 \end{equation}
Photomeson processes at the $\Delta$ resonance requires that
the Lorentz factor of the accelerated protons satisfies the resonance condition
$\gamma_f\sim\gamma_*$, which occurs near the peak of the thermal
photon distribution. One can show from (\ref{eq:eta}) that
for all three cases considered ($\epsilon=1/\theta^2_d$,
$\epsilon=1/\theta_d$, and $\epsilon=1$), the threshold 
$\gamma\geq\gamma_*(1-\Delta\varepsilon_0/\varepsilon_0)$ is satisfied. 
The pion energy is $\varepsilon_\pi\sim 0.2\varepsilon_p$, 
where $\varepsilon_p=\gamma_*\mu_p$ is the proton energy (in $m_ec^2$). Assuming that the
energy is evenly distributed among the four products and that pions decay before 
significant cooling or acceleration occurs, one may estimate the energy of neutrinos as
$\varepsilon_\nu=\varepsilon_\pi/4\sim0.05\varepsilon_p\approx
10^3\mu_p$, about 1 TeV.  

In a strong thermal radiation field, energy loss of charged pions due to inverse Compton scattering
can be important. The decay time is $t^D_\pi=2.6\times10^{-8}\gamma_\pi\,{\rm s}\approx
5\times10^{-4}\,\rm s$ for $\gamma_\pi=3.7\times10^4$ in the observer's frame.
The energy loss rate of pions is $-\dot{\gamma}_\pi\approx c\sigma_{T\pi} \varepsilon_{th}n_{ph}
\gamma^2_\pi(m_e/m_\pi)$, where the cross section is $\sigma_{T\pi}=(m_e/m_\pi)^2\sigma_T\approx 
1.3\times10^{-5}
\sigma_T$, $m_\pi\approx 274m_e$ is the pion mass (about 140 MeV; and the neutral pion mass
is about 135 MeV).  The cooling time can be obtained from
\begin{eqnarray}
t^{IC}_\pi&\approx& \left|{\gamma_\pi\over\dot{\gamma}_\pi}\right|
\nonumber\\
& \approx &10^{-6}\left({5\times10^7\,{\rm K}\over T_s}\right)^4\left({R_0
\over r}\right)^{-2}\left({3.7\times10^4\over\gamma_\pi}\right).
\end{eqnarray}
Since the threshold for pion production is reached before the proton reachs
$s_f$, charged pions produced by the proton can be accelerated and this
acceleration time is much shorter than either the cooling time and 
decay time. The pion has been accelerated to very high energy before
it decays. Therefore, the typical pion energy is
$\varepsilon_\pi=0.2\gamma_*\mu_p+(\gamma_f-\gamma_*)\mu_p=(\gamma_f-0.8\gamma_*)\mu_p
\sim 10^7$, which would give rise to the neutrino energy 
$\varepsilon_\nu\sim\varepsilon_\pi/4\sim 2.5\times10^6$ or 1.3 TeV.
However, in the region above $s_f$, the pion suffers energy loss due to IC and the
the neutrino energy can be much lower than the above estimate.
The condition $t^D_\pi=t^{IC}_\pi$ leads to $\gamma_\pi=1.2\times10^3(5\times10^7\,
{\rm K}/T_s)^2$ at $r\sim R_0$ and $\gamma_\pi\approx5.6\times10^3$ at $r\sim R_{LC}$,
corresponding to $\varepsilon_\nu=\varepsilon_\pi/4\approx 8.2\times10^4$ (or
40 GeV) and $3.8\times10^5$ (or $2\times10^2$ GeV).

The optical depth for photomeson production can be approximated by
$\tau_\pi\approx n_{\rm ph}\sigma_{\pi}R_0\Delta s$,
where $\sigma_{\pi}\approx 2\times10^{-28}\,{\rm cm}^2$ is the cross section at the $\Delta$ resonance
peak, $\Delta s$ is the characteristic length for pion production, and $n_{\rm ph}$ is
the photon number density given by (\ref{eq:nph1}).
Assuming $\tau_\pi=1$ one obtains $\Delta s\approx 1/n_{\rm ph}\sigma_{\pi}R_0\approx
0.12$ at $r\approx R_{LC}=4.7\times10^6\,\rm
cm$ and one would have a much smaller $\Delta s$ than this if $r\sim R_0$.
The neutrino emissivity is $A_\nu\sim (\varepsilon_\pi/4)\sigma_\pi
n_{ph}n_{GJ}c$ with $n_{GJ}\approx 3.5\times10^{16}
(B_0/10^{15}\,{\rm G})(P/1\,{\rm ms})^{-1}\,{\rm cm}^{-3}$, 
from which one may derive the neutrino luminosity 
$L_\nu\sim 4\pi(\theta_d R_0)^2R_0A_\nu$. The neutrino number flux 
density can be approximated by
\begin{eqnarray}
F_\nu&\approx& {L_\nu\over\Delta\Omega_0D^2\varepsilon_\nu}\approx 1.7\times10^{-11}
{1\over\Delta\Omega_0}\left({P\over1\,{\rm ms}}\right)^{-2}
\left({B_0\over10^{15}\,{\rm G}}\right) \nonumber\\
&&\times
\left({T_s\over5\times10^7\,{\rm K}}\right)^3
\left({D\over10\,{\rm Mpc}}\right)^{-2}\,{\rm cm}^{-2}\,{\rm s}^{-1}.
\label{eq:Fnu}
\end{eqnarray}
where $\Delta\Omega_0$ is the beaming angular width, $D$ is the distance 
to the source, and one assumes $\Delta s R_0\sigma_\pi n_{\rm ph}=1$.  
Since neutrinos are generally detected through production of muons, the event rate can
be calculated from $N_\nu\approx F_\nu P_{\nu\mu}$, where $P_{\nu\mu}=
10^{-7}(\varepsilon_\nu/2\times10^3)$, with $\varepsilon_\nu=2\times10^3$
corresponding to 1 GeV, is the probability of the muon event (Halzen \& Hooper 2002). 
Then, one finds
\begin{eqnarray}
N_\nu&\approx& 54{\tau_\pi\over \Delta\Omega_0}
\left({P\over1\,{\rm ms}}\right)^{-2}\left({B_0\over10^{15}\,{\rm G}}\right)
\left({T_s\over5\times10^7\,{\rm K}}\right)^3\nonumber\\
&&\times\left({D\over10\,{\rm Mpc}}\right)^{-2}
\left({\varepsilon_\nu\over 100\,{\rm GeV}}\right)
\,{\rm km}^{-2}\,{\rm yr}^{-1}.
\end{eqnarray}
The event rate in the early spin-down phase is about
$N_\nu=54\,{\rm km}^{-2}\,{\rm yr}^{-1}$ with the neutrino energy 
about 100 GeV. Such rate may be too low to be detectable with the current AMANDA 
and requires a km scale detector with a lower energy threshold 
of about 100 GeV.

The photomeson process also produces relativistic neutrons, which move ballistically through the 
magnetosphere and may interact with target nuclei in the stellar envelope to produce neutrinos and 
gamma-rays (Protheroe, Bednarek \& Luo 1998). The neutron decay time is
$t_n\sim (920\,{\rm s})\gamma_n$, where $\gamma_n$ is the Lorentz factor 
of neutrons. For $\gamma_n\sim \gamma_f\sim 5\times10^4$, the decay distance
is about $t_nc\sim 1.3\times10^{18}\,\rm cm$, much larger than the size
of the shell $\sim v_st$, where $v_s\sim 3000\,{\rm km}\,{\rm
s}^{-1}$ is the expansion speed. If the shell is completely fragmented
(Arons 2003), neutrons would propagate freely through it and decay outside 
the remnant.

\section{Proton acceleration in the outer magnetosphere}

Protons may be accelerated to much higher energy near the LC where
screening of the electric field occurs in the perpendicular (to the field
line) direction and the maximum energy is limited only by radiation loss.
To accelerate protons in the outer magnetosphere, a specific mechanism for 
injection of protons into the gap is needed and will be discussed in Sec.~4.1.

\subsection{MSM outer gap}

In close analogy to the outer gap in normal young pulsars, we assume an oblique
rotator and that an outer gap develops in a region from the null surface along the open 
field lines to the LC, with an upper boundary surface defined by a pair production front 
and a lower boundary by the last open field lines. As in the case of the polar gap,
such gap can be nonstationary due to the very nature of time-dependent pair cascades.
The gap energetics is thus determined by the average thickness (between the two boundary 
surfaces) $w_g$ of the gap, where $w_g$ is determined as in the
PC scenario by the pair production free path 
of an electron or a positron (see discussion below Eq 24). 
Since the available potential $w_g\phi_m$ well exceeds
that for radiation-reaction limit of protons due to curvature radiation, the oscillatory
nature of the gap can cause modulation of the particle flux and it does not 
constrain the proton's maximum energy (it is limited by radiation reaction). 
One may estimate the maximum proton energy and relevant particle flux by 
following similar procedure to that used in the steady outer gap model
(e.g. Cheng, Ho, Ruderman 1986; Romani 1996; Hirotani, Harding \& Shibata
2003) and the result should not depend on particular aspects of the model.

\begin{figure}
\psfig{file=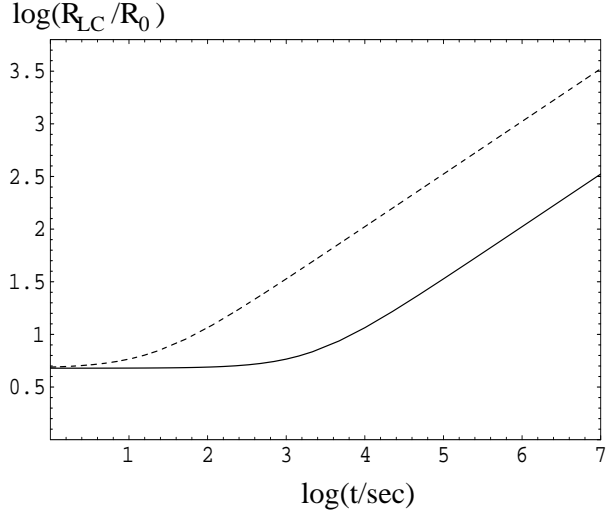,width=8cm}
\caption{Evolution of the light cylinder radius. The solid and dashed 
lines correspond respectively to $B_0=10^{15}\,\rm G$ and
$10^{16}\,{\rm G}$.
}
\label{fig:rlc}
\end{figure}

\begin{figure}
\psfig{file=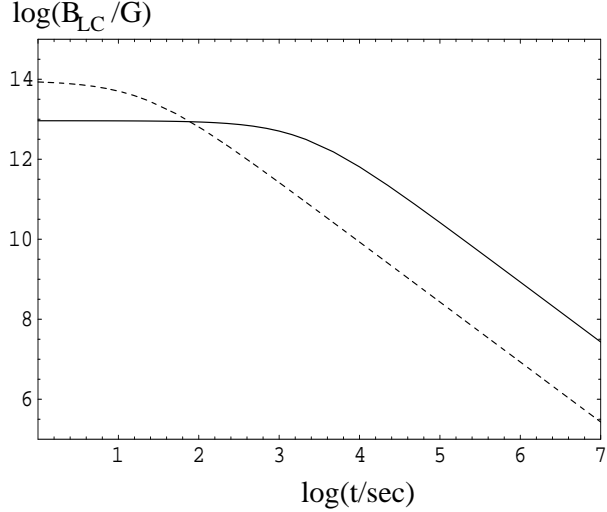,width=8cm}
\caption{Evolution of the magnetic field $B_{LC}$ at the
light cylinder radius. The solid and dashed
lines correspond respectively to $B_0=10^{15}\,\rm G$ and
$10^{16}\,{\rm G}$.
}
\label{fig:blc}
\end{figure}

It is assumed that particles are accelerated along the field lines to ultra-high 
energies and emit high energy photons through curvature radiation or
inverse Compton scattering and these photons convert into pairs
through single photon decay in the strong magnetic field. Note that
in normal young pulsars the only effective channel for pair creation near
the LC is through photon-photon collisions. Since the field lines curve away from the 
magnetic axis, the pairs are produced on upper, neighbouring field lines and 
thus one may estimate $w_g$ in a way similar to that 
for the polar gap, i.e. by estimating the pair production free path of
an electron (or a positron). Because the magnetic field near the LC is weaker than 
near the PC, the pair production free path should be calculated from 
the opacity of photon absorption (due to pair production), characterized by a parameter
$\chi=(1/2)\varepsilon_\gamma\varepsilon_B\sin\theta_{kB}$, where
$\varepsilon_\gamma$ is the energy of pair producing photons and $\theta_{kB}$
is the photon propagation angle as defined in (\ref{eq:threshold1}) and
(\ref{eq:threshold2}).
Apart from the threshold conditions (\ref{eq:threshold1})
and (\ref{eq:threshold2}), the following condition is needed for producing 
one pair (corresponding to the opacity $\tau\sim1$)
(e.g. Arons 1983):
\begin{equation}
\chi\geq {1\over15}\left({20\over\ln\Lambda}\right),
\end{equation} 
where the near-threshold effect on the opacity (Daugherty \& Harding 1983)
is ignored and $\ln\Lambda$ is a logarithmic parameter 
defined through $\Lambda(\ln\Lambda)^3=
1.6(\alpha_f/\varepsilon^2_\gamma\varepsilon_B)(R_c/\lambda_c)$, which is not sensitive to
the pulsar parameters, with a value between 1 and 30. For $\varepsilon_\gamma=5\times
10^6$, one has $\ln\Lambda\sim2$. For a photon of a given energy
$\varepsilon_\gamma$, the typical angle at which the photon converts into
a pair is 
\begin{equation}
\sin\theta_{kB}\approx 1.2\times10^{-6}
\xi^3\left({\varepsilon_\gamma\over5\times10^6}\right)^{-1}
\left({P\over 1\,{\rm ms}}\right)^3
\left({B_0\over10^{15}\,{\rm G}}\right)^{-1},
\end{equation}
where $\xi=r/R_{LC}$.
The characteristic thickness (in $R_{LC}$) of the gap can be estimated from 
$w_g\sim \sin\theta_{kB}$. 
The time dependence of the LC radius and the magnetic field
at the LC are shown in Figures~\ref{fig:rlc} and \ref{fig:blc}, 
respectively. Single photon decay should be the dominant 
pair production process up to $t\sim 10^{6.8}\,\rm s$ when the magnetic field
at the LC drops below $10^8\,\rm G$. When $P$ is sufficiently long, magnetic 
pair production is no longer important and the only channel for pair production 
is through photon-photon collisions and like normal young pulsars the MSM becomes
more and more efficient with increasing $P$ to a maximum where the outer gap is completely open.

One problem with acceleration of protons in the outer magnetospheric
region is the proton injection. An outer gap model including 
injection of particles was recently discussed in detail by Hirotani 
\& Shibata (1999), also Hirotani, Harding 
\& Shibata (2003). (In their discussions, protons are not included in the model.) 
It has been recognized that current circulation       
in the pulsar-wind system may play an essential role in transforming
rotational energy to particle kinetic energy (e.g. Shibata
1991). Such global current should exist even in the case of
nonstationary acceleration. Thus, one may assume that the ion injection is 
due to the global current system. The sign of the accelerating electric field 
is determined by the sign of $B_z$. For an outer gap located outside the null surface
($B_z=0$), the accelerating field is positively directed
when $B_z<0$, corresponding to pulsars with $\mbox{\boldmath $\mu$}_m\cdot{\bf \Omega}>0$.
Therefore,  outflowing primary electrons from the PC and downflowing positrons (possible protons) 
form a return current directed toward the PC, and a flow of protons plus positrons
through the outer gap forms part of the outward-directed current.
In the steady model, inclusion of an external flux shifts the location of the
zero GJ density, the `null surface', and hence the gap location. For a flux injected 
from the null surface, the zero GJ density is shifted towards to the LC, which 
shifts the gap to the LC (Hirotani, Harding \& Shibata 2003). Although such 
features are derived based on the steady assumption, they may well be applicable
to the nonstationary case if acceleration confined to the region very 
close to the last open field lines where the influence of
nonstationary pair cascades is the least.

\subsection{Maximum proton energy}

The outer gap potential, $\phi_mw_g$, limited by pair production by positrons/electrons,   
is much larger than that of the polar gap, $\phi_ms^2_f$. For example, at $P=1\,\rm ms$,
the outer gap has $10^{-6}\phi_m$ ($w_g\sim10^{-6}$ at $P\sim 1\,\rm ms$), 
as compared to the polar gap
$10^{-8}\phi_m$ ($s_f\sim 10^{-4}$, cf. Eq.~\ref{eq:sf2}). For the relevant parameter regime 
one is interested in,  pair cascades do not constrain the acceleration length
since pairs are produced on field lines with a smaller magnetic colatitude. 
So, the maximum proton energy is limited only by radiation
reaction such as curvature radiation. Since the specific form of the potential is model 
dependent, here we write the parallel accelerating electric field in the form 
\begin{equation}
E_\Vert={\phi_mf_aw_g\over 2R_{LC}}\xi^\alpha,
\end{equation}
where $f_a<1$ is the fraction of the full (vacuum) potential across the gap, 
$r_{NS}/R_{LC}\leq \xi\leq 1$, 
$r_{NS}\approx R_{LC}(2\cot\alpha_i/3)^{2}$ (in the $\Omega$-$B$ plane)
is the radial distance to the null surface, $\alpha_i$ is the MSM inclination 
angle (relative to the spin axis), and $\phi_m$ is given by
(\ref{eq:fim}). The power index is in the range of $-3\leq\alpha\leq0$.  In Romani (1996)'s 
model, $\alpha=-1$ is assumed. For an oblique rotator, $r_{NS}$ is a 
substantial fraction of the LC radius, one may use an approximation $\xi\sim1$. 

The maximum proton Lorentz factor can be derived by equating
the energy loss rate due to curvature radiation to the acceleration rate. 
Note that the power of curvature emission for protons is the same as that for 
electrons, given by (\ref{eq:Pc}) with $\gamma$ now being interpreted as
the proton Lorentz factor. The radiation-reaction limited Lorentz factor
is derived as
\begin{eqnarray}
\gamma_c&\approx& \left({3\gamma_mf_aR_{LC}\over4r_e}
\right)^{1/4}\left({w_g\over\xi^{3/2}}\right)^{1/4}\xi^{(2\alpha+5)/8}\nonumber\\
&=&7.6\times10^8\,f^{1/4}_a
 \left({P\over1\,{\rm ms}}\right)^{-1/4}\left({B_0\over10^{15}\,{\rm G}}\right)^{1/4}
\left({w_g\over\xi^{3/2}}\right)^{1/4}\xi^{(2\alpha+5)/8}.
\label{eq:gc1}
\end{eqnarray}
For $f_a=10^{-2}$, one has $\gamma_c\sim 2\times10^8(w_g/\xi^{3/2})^{1/4}
(P/1\,{\rm ms})^{1/4}$.
The radius of the open field line region can be written as
$R_\perp\sim\xi^{3/2}R_{LC}$.
The condition $w_g\lapprox R_\perp/R_{LC}$ implies that $w_g/\xi^{3/2}\lapprox 1$. 
As one is interested only in the outer gap near the LC, $\xi\sim1$, 
the radiation-reaction limited Lorentz factor is not sensitive to the specific
choice of $\alpha$ value. 

\section{Photomeson production near the LC}

In the outer magnetospheric region, both thermal radiation from
the star's surface and nonthermal radiation from cascades due to 
particle acceleration in the outer gap can provide target photons
for photomeson processes. Accelerated protons may also interact
with soft photons from the diffuse thermal radiation, e.g. from the
hot ejecta or the interaction region between the MSM wind and the remnant
shell. Here, we consider the first and third possibilities only.

\subsection{Photomeson threshold}

The thermal radiation from the star's surface subtends a much smaller
solid angle near the LC than near the PC and thus the protons satisfy the photomeson
threshold at much higher energies. Assuming that the acceleration occurs near 
the last open field lines, the characteristic propagation angle of thermal photons 
originating from the star's surface can be derived from
\begin{equation}
\cos\theta_m\approx 1-{\theta^4_d\over2\xi^2},
\end{equation}
where $\theta^4_d/\xi^2\ll1$.
Since this condition is generally satisfied, the geometric effect of the 
thermal radiation is important, which leads to the following expression
for the photomeson production threshold for the photon energy,
\begin{equation}
\varepsilon_{ph}\approx
0.35{\mu_p\xi^2\over\gamma\theta^4_d}.
\label{eq:eph}
\end{equation}
Since $\xi^2/\theta^4_d\gg1$, protons satisfy the threshold at much higher energies
than near the PC, producing pions with extremely high energy. Eq. (\ref{eq:eph}) 
implies that $\varepsilon_{ph}\sim P^2$, higher proton energy is required as the MSM
spins down to a longer period. At $P\sim 50\,\rm ms$, one has $\gamma\geq10^8$.

Charged pions are subject to the energy loss due to IC. Since the Lorentz factor of
pions is $\gamma_\pi\gapprox 0.2\gamma_*(m_p/m_\pi)$, the photon energy in the
pion rest frame is $\varepsilon'\sim
2.8\Theta\gamma_\pi\xi^2/2\theta^4_d\gapprox1.3\times10^4
(m_p/m_\pi)(5\times10^7\,{\rm K}/T_s)^{-1}\gg1$. In the following discussion,
the surface temperature $T_s$ is assumed to be time independent.
Hence, IC is in the KN regime with a characteristic time:
\begin{eqnarray}
t^{KN}_\pi&\approx&
{120\lambda^3_c\gamma_\pi\mu_\pi\xi^2
\over\pi^2\sigma_Tc\Theta^2\theta^4_d\ln(4\gamma_\pi\Theta)}
\nonumber\\
&\approx& (1.4\times10^{-10}\,{\rm s})\left({T_s\over5\times10^7\,{\rm K}}\right)^{-2}
{\gamma_\pi\xi^2\over \theta^4_d\ln(4\gamma_\pi\Theta)}.
\end{eqnarray}
We assume that the thermal radiation field originating from the neutron star's
surface has spherical symmetry with the photon number density decreasing radially
as $1/r^2$. The LC radius increases as the MSM spins down (cf. Figure~\ref{fig:rlc}).  
Therefore, the photon number density near the LC reduces as the magnetosphere 
expands. One may find the characteristic period $P_1$ at which the energy loss time of 
charged pions near the LC is equal to the pion decay time. This period is given by
\begin{equation}
P_1\sim (11.8\,{\rm ms})(T_s/5\times10^7\,{\rm K}){[\ln(4\gamma_\pi\Theta)]^{1/2}\over\xi}.
\end{equation}
Charged pions lose much of their energy through IC until the MSM spins down to
$P\sim P_1\approx21\,\rm ms$ when the photon number density is sufficiently low
and IC can no longer constrain the neutrino energy. The neutrino energy can reach 
$\varepsilon_\nu\sim 0.05\gamma_c\mu_p=5\times10^9$, about $10^3$ TeV.

If the region surrounding the MSM is filled with diffuse thermal radiation,
extremely energetic protons produce pions on these soft photons.  There are two possible
sources for the diffuse thermal radiation: the hot ejecta in the early 
phase of the pulsar spin-down (e.g. Beall \&
Bednarek 2002), and the interaction region of the inner surface of the remnant
where the MSM wind may deposit its energy (e.g. Rees \& M\'esz\'aros 2000).
The threshold condition for photopion production is
$\gamma\sim 3.2\times10^6(5\times10^5\,{\rm K}/T_d)$,
where $T_d$ is the temperature of the thermal radiation. Here we assume that 
the radiation field is isotropic. This condition is well satisfied 
by (\ref{eq:gc1}). For such a low $T_d$, pion cooling is not important
in constraining the neutrino energy.  

\subsection{Neutrino flux}

The neutrino flux can be estimated from the proton luminosity, given by
\begin{eqnarray} 
L_\nu&\sim & {w_g \over4\xi^{3/2}}\,2\pi\theta^2_dR^2_0n_{GJ}cm_pc^2\gamma_c f_p
\nonumber\\
&\approx&
(10^{45}\,{\rm erg}\,{\rm s}^{-1})\,
\left({P\over1\,{\rm ms}}\right)^{-2}\left({B_0\over10^{15}\,{\rm G}}\right)\,
{w_gf_p\over\xi^{3/2}},
\end{eqnarray}
where the proton flux in the gap is assumed to be a fraction $f_p\leq1$  of
the GJ flux $n_{GJ}c$. The neutrino number flux density can be estimated from 
$F_\nu\sim L_\nu/4\pi\Delta\Omega_0D^2\varepsilon_\nu\approx
(4.3\times10^{-12}\,{\rm cm}^{-2}\,{\rm s}^{-1})(P/1\,{\rm
ms})^{-2}\,(f_pw_g/\xi^{3/2})$ for $\Delta\Omega_0=4\pi$, 
$B_0=10^{15}\,{\rm G}$ and $D= 10\,{\rm Mpc}$. 
The atmospheric background neutrinos have the flux spectrum
$\phi^b_\nu\sim 10^{-7}(\varepsilon_\nu/2\times10^6)^{-2.5}\,{\rm cm}^{-2}\,
{\rm s}^{-1}\,{\rm sr}^{-1}$. The planned IceCube has an angular resolution
of about $1^\circ$. The background  number flux density from a small patch
$(\pi/180)^2$ of the sky is $F^b_\nu\sim\phi^b_\nu(\pi/180)^2
\sim 9.6\times10^{-19}\,{\rm cm}^{-2}\,{\rm s}^{-1}$ at $\varepsilon_\nu\sim 
2\times10^9$ or about $10^3\,\rm TeV$. The predicted flux is well above 
this background flux for $f_p>10^{-6}$.

One estimates the maximum neutrino event rate as
$N_\nu\sim F_\nu P_{\nu\mu}$ with $P_{\nu\mu}\sim 2\times10^{-6}(\varepsilon_\nu/
2\times10^6)$ for neutrinos above TeV energies (Halzen \& Hooper 
2002), leading to
\begin{equation}
N_\nu \sim 
1.8\times10^4f_p\left({P\over1\,{\rm ms}}\right)^{-2}
\left({w_g\over\Delta\Omega_0\xi^{3/2}}\right)\left(
{D\over10\,{\rm Mpc}}\right)^{-2}\,{\rm km}^{-2}\,{\rm yr}^{-1}
\end{equation}
where one assumes $B_0=10^{15}\,{\rm G}$. For example, for
$P\sim 48\,\rm ms$, for which one has $w_g/\xi^{3/2}\sim0.13$, one
has an event rate of about $N_\nu\sim (1.1\,{\rm km}^{-2}\,{\rm yr}^{-1})
(f_p/\Delta\Omega_0)$, which is potentially detectable by the IceCube provided 
that the proton flux in the gap is a substantial fraction of the GJ flux and 
that the emission is moderately beamed, say $\Delta\Omega_0\sim1$ and the source is relatively
nearby. Since the specific value of $f_p\leq1$ is not well constrained, one cannot exclude
the possibility that the current that forms a closed circuit is predominantly due to positrons. 
In this case, the neutrino flux produced from the outer gap would be strongly limited
by the factor $f_p\ll1$.

\section{Discussion and conclusions}

Neutrino production in the magnetosphere of a fast rotating magnetar is considered. 
It is shown that MSM can be a strong source of high energy neutrinos in its
early spin-down phase. We consider proton acceleration in both the inner
magnetosphere near the PC and the outer magnetosphere near the LC. 
When protons are accelerated near the PC,
their maximum energy is constrained by pair production due to accelerated seed
positrons/electrons or primary protons.  In supercritical magnetic fields, photon splitting 
may be important in limiting pair production (Baring \& Harding 2001). However,
since this process can only split the $\perp$ polarized photons (Baring \& Harding 2001; 
Usov 2002), it may not prohibit pair production by the $\Vert$ photons. 
Pair annihilation is greatly enhanced in a supercritical magnetic field
leading to reduction in the secondary pair density, but such process
does not strongly affect the electric field screening. Due to the supercritical 
magnetic field, the efficiency of neutrino production can be estimated 
from the pair production free paths of positrons/electrons. The 
gap determined by pair production by protons has a relatively higher
efficiency (cf. Figure 3). Although the proton-controlled steady gap cannot
be excluded, the assumption that electrons/positrons control the dynamics of the 
polar gap appears to be the more consistent with the lack of any observational 
evidence for two classes of radio pulsars. If protons play a role in determining the polar gap,
one would have one distinct class of pulsars with their gap controlled by
electrons/positrons and the other by protons. For a typical radio pulsar with
$T_s\sim 10^6\,\rm K$, the proton-controlled gap is very inefficient in producing
pair plasmas that  needed for production of the observed coherent radio emission.
So far, there is no obvious observational evidence in pulsar radio emission for 
this distinction. In the case of the gap limited by pair production by 
electrons/positrons, photomeson production is possible provided 
that the surface temperature is $T_s\sim 5\times10^7\,\rm K$. 
Because of the limit on the proton energy by pair production, only a tiny fraction 
(about $\eta\sim 10^{-8}$ initially) of the spin-down power goes into protons and
efficiency increases as the MSM spins down while the proton flux decreases. 
As photomeson production occurs near the PC, charged pions are subject to 
energy loss to inverse Compton scattering, which further limits the neutrino 
energy to about 100 GeV, below the threshold for IceCube.

Acceleration in the outer gap can be more efficient than in the 
polar gap as the maximum proton energy, limited by curvature emission, can 
reach about $10^8m_pc^2$. It is suggested here that proton injection in the outer gap
is part of a closed global current that flows through the gap in the form 
of proton flux. In the application to MSMs with $\mbox{\boldmath $\mu$}_m\cdot{\bf
\Omega}>0$, the outflowing particles in the PC region
are mainly electrons that, together with possible downflowing positrons or protons,
provide a return current into the PC. The outer gap is
limited by pair production due to magnetic single photon decay. The efficiency is 
initially small due to efficient pair production limiting the gap thickness,
which in turn limits the total flux of protons that can pass through and
accelerated in the gap. The efficiency increases with increasing period.
Since the thermal radiation from the surface subtends a much 
smaller solid angle, protons satisfy the photomeson threshold at much higher
energies so that pions are produced at high energies $\sim 10$ TeV. 
As the star spins down rapidly, the magnetosphere expands and the photon
number density in the outer magnetosphere decreases. When $P\sim P_1$, charged 
pions decay into neutrinos before they lose their energy and TeV neutrinos emerge.
TeV neutrinos can also be produced as a result of interactions
of ultrarelativistic protons originating from the magnetosphere with
soft photons from the diffuse thermal radiation due to the hot ejecta,
which may exist in the early phase of a new born MSM, or due to the heating 
of the interface region between the relativistic MSM wind and the
remnant shell. The neutrino number flux density is estimated and it is shown
that the corresponding muon event rate may be detectable with the planned
IceCube provided that the source is relatively nearby or the emission
is moderately beamed. The estimate is based on the assumption
that the outer gap is time-independent, which may not be realistic.
Inclusion of time-dependence of pair production and hence the
nonstationary gap may lead to a significant modification to
the particle injection and acceleration. A nonstationary gap does not have
a well defined pair production front and hence a much larger effective cross
section area, allowing more protons to be accelerated leading to an 
increase in the neutrino luminosity.   
In the above discussion, one ignores the possibility that
a positron flux forms a major part of the current that closes the 
current circuit (e.g. Shibata 1991), with a proton flux limited to a 
small fraction of the GJ flux. If this is the case, the neutrino flux 
is severely constrained by the fraction factor $f_p\ll1$.

So far, the polar gap and outer gap scenarios have been treated 
separately. In realistic situations, pair cascades near the polar 
cap may affect the outer gap and vice versa. To determine 
self-consistently such complication requires a quantitative model 
that links the two acceleration regions and such
model is currently not available. Apart from the polar gap and outer gap 
scenarios, protons may be accelerated to ultra-high energy in a 
slot gap (Arons 1983; Muslimov \& Harding 2004). In such model, protons 
can be assumed to be primary particles extracted from the
PC and can be accelerated to energies well exceeding the pion
production threshold (e.g. Protheroe, Bednarek, \& Luo 1998).
Since a slot gap may accelerate protons to the maximum energy limited
by radiation-reaction and the available particle flux (through the gap)
is constrained by pair production, the result (the neutrino energy and flux) 
discussed in Sec. 5 should be qualitatively valid for this case as well. 
However, a quantitative prediction of the neutrino flux from the slot gap requires 
further work. 

\section*{Acknowledgement}

The author thanks Don Melrose for helpful comments on the manuscript.


\begin{thebibliography}{22}
\bibitem{a83}
Arons, J., 1983, ApJ, 266, 215.
\bibitem{a03}
Arons, J., 2003, ApJ, 589, 871.
\bibitem{akp78}
Asseo, E., Kennel, C. F., Pellat, R., 1978, A\&A, 65, 401.
\bibitem{bh92}
Baring, M., Harding, A., 1992, in Compton Workshop, 245.
\bibitem{bh01}
Baring, M., Harding, A., 2001, ApJ, 547, 929.
\bibitem{bb02}
Beall, J. H., Bednarek, W., 2002, ApJ, 569, 343.
\bibitem{bp99}
Bednarek, W., Protheroe, R., 1997, Phys. Rev. Lett., 79, 2616.
\bibitem{bp02}
Bednarek, W., Protheroe, R., 2002, Astroparticle Phys. 16, 397.
\bibitem{bp78}
Berezinsky, V. S., Prilutsky, O. F., 1978, A\&A, 66, 325.
\bibitem{by98}
Blackman, E. G., Yi, I., 1998, ApJ, 498, L31.
\bibitem{chr86}
Cheng, K. S., Ho, C., Ruderman, M., 1986, ApJ, 300, 500.
\bibitem{czs92}
Chodorowski, M., Zdziarski, A., Sikora, M., 1992, ApJ, 400, 181. 
\bibitem{detal92}
Dar, A., Kozlovsky, B., Nussinov, S., Ramaty,
R., 1992, ApJ, 388, 164.
\bibitem{db80}
Daugherty, J. K., Bussard, R. W., 1980, ApJ, 238, 296.
\bibitem{dh83}
Daugherty, J. K., Harding, A. K., 1983, ApJ, 273, 761.
\bibitem{dt92}
Duncan, R., Thompson, C., 1992, ApJ, 392, L9.
\bibitem{getal02}
Gaensler, B. M., et al., 2002, ApJ, 569, 878.
\bibitem{getal04}
Gaensler, B. M., et al., 2004, ApJ, in press.
\bibitem{g-etal00}
Gonthier, P., et al., 2000, ApJ, 540, 907.
\bibitem{gg03}
Granot, J., Guetta, A., 2003, Phys. Rev. Lett. 90, 191102.
\bibitem{go71}
Gunn, J. E.,  Ostriker, J. P., 1971, ApJ, 166, 523.
\bibitem{hh02}
Halzen, F., Hooper, D., 2002, Rep. Prog. Phys. 65, 1025.
\bibitem{hbg97}
Harding, A., Baring, M., Gonthier, P., 1997, ApJ, 476, 246.
\bibitem{hm98}
Harding, A., Muslimov, A., 1998, ApJ, 508, 328.
\bibitem{hm01}
Harding, A.,  Muslimov, A., 2001, ApJ, 
\bibitem{h79}
Herold, H., 1979, Phys. Rev. D19, 2868.
\bibitem{ha01}
Hibschman, J., Arons, J., 2001, ApJ, 554, 624.
\bibitem{hhs03}
Hirotani, K., Harding, A., Shibata, S., 2003, ApJ, 591, 334.
\bibitem{hs99}
Hirotani, K., Shibata, S., 1999, MNRAS, 308, 54.
\bibitem{k05}
Kirk, J., 2005, astroph/0504410.
\bibitem{l01}
Lai, D., 2001, Rev. Mod. Phys. 73, 629.
\bibitem{l04}
Luo, Q., 2005, PASA, 22, in press.
\bibitem{lk01}
Lyubarsky, Y., Kirk, J., 2001, ApJ, 547, 437.
\bibitem{mm96}
Melatos, A., Melrose, D.B., 1996, MNRAS, 279, 1168.
\bibitem{mh04}
Muslimov, A., Harding, A., 2004, ApJ, 606, 1143.
\bibitem{n04}
Nagataki, S., 2004, ApJ, 600, 883
\bibitem{pbl98}
Protheroe, R. J., Bednarek, W., Luo, Q., 1998, Astrop. Phys. 9, 1.
\bibitem{rm00}
Rees, M., M\'esz\'aros, P., 2000, ApJ, 545, L73.
\bibitem{r96}
Romani, R., 1996, ApJ, 470, 469.
\bibitem{rs75}
Ruderman, M., Sutherland, P. G., 1975, ApJ, 196, 51.
\bibitem{s78}
Sato, H., 1978, Prog. Theor. Phys. 58, 549.
\bibitem{s91}
Shibata, S., 1991, ApJ, 378, 239.
\bibitem{smt98}
Shibata, S., Miyazaki, J., Takahara, F., 1998, MNRAS, 295, L53.
\bibitem{s-etal91}
Stecker, F., Done, C., Salamon, M. H., Sommers, P., 1991, Phys. Rev. Lett. 66, 2697.
\bibitem{td95}
Thompson, C., Duncan, R., 1995, MNRAS, 275, 255.
\bibitem{u02}
Usov, V., 2002, ApJ, 572, L87.
\bibitem{u94}
Usov, V., 1994, MNRAS, 267, 1035.
\bibitem{u96}
Usov, V., Melrose, D.B., 1996, ApJ, 464, 306.
\bibitem{wb97}
Waxman, E., Bahcall, J. N., 1997, Phys, Rev. Lett., 78, 2292.
\bibitem{wm02}
Weise, J., Melrose, D. B., 2002, MNRAS, 329, 115.
\bibitem{w-etal00}
Wheeler, J. C., Yi, I., H\"oflich, P., Wang, L., 2000, ApJ, 537, 810.
\bibitem{wt04}
Wood, P. M., Thompson, C., 2004, astroph/0406133. 
\bibitem{zh-etal03}
Zhang, B., Dai, Z. G., M\'esz\'aros, P., Waxman, E., Harding, A. K., 2003, ApJ, 595, 346

\end{thebibliography}
\end{document}